\documentclass[%
 aip,
rsi,%
 amsmath,amssymb,
 reprint,%
]{revtex4-1}

\usepackage{graphicx}
\usepackage{dcolumn}
\usepackage{bm}

\begin{document}

\preprint{AIP/123-QED}

\title{Tunable anisotropic absorption in monolayer black phosphorus using critical coupling}
\author{Tingting Liu}
 \affiliation{Laboratory of Millimeter Wave and Terahertz Technology, School of Physics and Electronics Information, Hubei University of Education, Wuhan 430205, China}  
\author{Xiaoyun Jiang}
\affiliation{Wuhan National Laboratory for Optoelectronics, Huazhong University of Science and Technology, Wuhan 430074, China}  
\author{Huaixing Wang}
\affiliation{Laboratory of Millimeter Wave and Terahertz Technology, School of Physics and Electronics Information, Hubei University of Education, Wuhan 430205, China}  
\author{Yong Liu}
\affiliation{Laboratory of Millimeter Wave and Terahertz Technology, School of Physics and Electronics Information, Hubei University of Education, Wuhan 430205, China}  
\author{Chaobiao Zhou}
\affiliation{College of Mechanical and electrical engineering, Guizhou Minzu University, Guiyang 550025, China}  
\author{Shuyuan Xiao}%
 \email{syxiao@ncu.edu.cn}
\affiliation{Institute for Advanced Study, Nanchang University, Nanchang 330031,China
}%

\begin{abstract}
We present a monolayer black phosphorus (BP)-based metamaterial structure for tunable anisotropic absorption in the mid-infrared. Based on the critical coupling mechanism of guided resonance, the structure realizes the high absorption efficiency of 99.65$\%$ for TM polarization, while only 2.61$\%$ at the same wavelength for TE polarization due to the intrinsic anisotropy of BP. The absorption characteristics can be flexibly controlled by changing critical coupling conditions, including the electron doping of BP, geometric parameters and incident angles of light. The results show feasibility in designing high-performance BP-based optoelectronic devices with spectral tunability and polarization selectivity.
\end{abstract}

\maketitle

Over the last decade, atomically thin two-dimensional (2D) materials with distinctive electronic and optical properties have stimulated revolutionary developments in the fields of optics and optoelectronics.\cite{xia2014two}  Black phosphorus (BP) with many unique properties not found in other members of 2D material family, has captured the interest of research community since 2014. \cite{li2014black} BP thin film shows a carrier mobility up to 1000 cm$^2$ V$^{-1}$ s$^{-1}$, indicating its promising application in high-frequency and high-speed optoelectronic devices. \cite{xia2014rediscovering,liu2014phosphorene} BP also has a moderate direct electronic band gap that is tunable from 0.3 eV to 2 eV, which is considered to bridge the energy gap between that of graphene and transition metal dichalcogenides.\cite{rodin2014strain,tran2014layer,deng2017efficient} Moreover, BP exhibits the high in-plane anisotropy in electronic and optical properties due to its puckered orthorhombic structure, offering opportunities for the design of polarization dependent devices.\cite{fei2014strain,qiao2014high} However, the optical absorption of BP thin film is quite low due to the nature of atomic thickness, setting an obstacle for practical applications. 

Considerable efforts have been devoted to enhance the light-BP interaction based on various schemes and metamaterial structures in the terahertz and infrared regime. The plasmonic resonance in nanostructured monolayer BP was widely studied for absorption enhancement whereas most of the resonance structures showed very weak absorption efficiency.\cite{low2014plasmons,liu2016localized,lu2017strong} In further works, the monolayer BP was integrated with different multilayer structures for the perfect absorption, such as within Fabry-Perot cavity, in a hyperbolic metamaterial or using the concept of coherent perfect absorption.\cite{wang2017dual,song2018biaxial,fo2018anisotropic,guo2019tunable,xiao2019tunable,zhu2019tunable,cai2019anisotropic} These multilayer structures brought another problem on the complicated fabrication and additional modulation configuration. Most recently, another perfect absorption mechanism based on critical coupled mode theory (CMT) was investigated in metamaterial structures, showing the advantages of the simple design and high absorption efficiency.\cite{piper2014total,zand2018multispectral,luo2018tunable,jiang2018approaching,li2018wavelength,qing2018tailoring} Due to these desirable capabilities, the perfect absorption in BP-based metamaterials via critical coupling is in urgent need.

In this letter, we design a perfect absorption metamaterial structure in mid-infrared by critically coupling monolayer BP with guided resonance in a periodic polymer structure. The proposed absorption structure not only exhibits different absorption characteristics under TM and TE polarizations originating from the in-plane anisotropy of BP, but also demonstrates the actively tunable absorption by manipulating the critical coupling conditions, such as electron doping of BP, structural parameters and incident angles of light. These properties of the structure may allow for the development of high-performance BP-based optoelectronic devices that will take advantage of the dynamic tunability and polarization dependence.

\begin{figure} [htbp]
	\label{figure1}
	\includegraphics
	[scale=0.4]{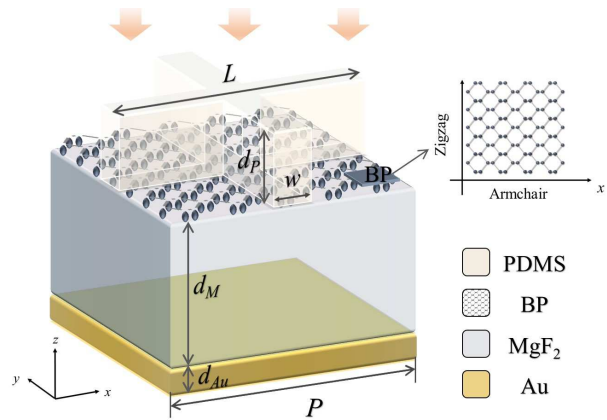} 
	\caption{\label{fig:FIG1}{Schematic of one unit cell of the proposed absorption structure. The structure consists of orthogonally crossed PDMS bars, monolayer BP, MgF$_2$ layer and Au mirror from top to bottom. The inset represents the in-plane anisotropy of BP along the $x$-direction (armchair direction) and $y$-direction (zigzag direction) due to the puckered arrangement of BP atoms.
			}}	

\end{figure}

The proposed absorption structure is schematically shown in Fig. 1. The monolayer BP is placed between the orthogonally crossed Polydimethylsiloxane (PDMS) array and the MgF$_2$ layer, and the sandwiched structure is mounted on the top of a gold (Au) mirror. The refractive indices of the PDMS and MgF$_2$ are 1.37 and 1.34, respectively,\cite{dodge1984refractive,querry1987optical} and the low refractive index contrast greatly simplifies the fabrication processes in practice. The atoms in monolayer BP form a puckered hexagonal honeycomb structure, resulting in the in-plane anisotropy of BP along the x-direction (armchair direction) and y-direction (zigzag direction). With the reasonable assumption of thickness of BP $d_{BP}$=1 nm, the complex permittivity of BP is derived from the surface conductivity by a semiclassical Drude model and has direction-dependent properties along $x$ and $y$ polarization within the wavelength of interest.\cite{liu2016localized,xiong2017strong} The permittivity of Au mirror is calculated from Drude model.\cite{ordal1985optical} The numerical simulations in the work are conducted using finite-difference time-domain (FDTD) method. The plane waves are normally incident from z direction. Periodic boundary conditions are applied in $x$ and $y$ directions and perfectly matched layers are used along the propagation direction. Using Au mirror with a thickness of 200 nm to block the transmission, the absorption is calculated by $A=1-R$, where $R$ represents reflection. 

\begin{figure} [htbp]
	\label{figure2}
	\includegraphics
	[scale=0.45] {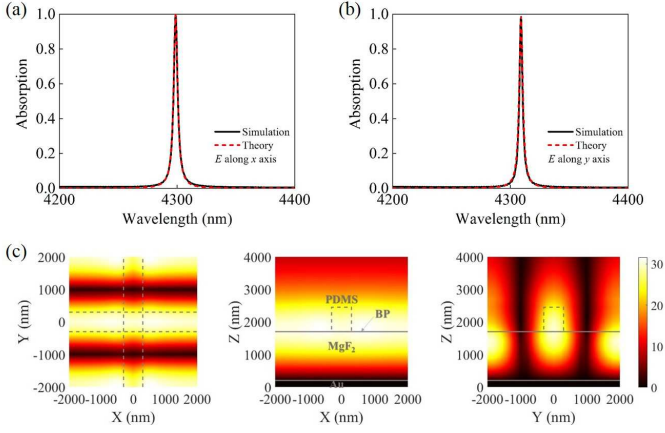}
	\caption{\label{fig:FIG2}{Comparisons between the simulated (solid lines) and theoretical (dotted lines) absorption spectra of the proposed BP-based structure for (a)TM polarization and (b) TE polarization. (c)Distributions of the electric fields $|E|$ of the structure at the resonance wavelength of 4304.27 nm for TM polarization. 			
	}}	
	
\end{figure}

To investigate the absorption characteristics of the proposed structure, the structural parameters are initially set as following. The period of the structure is $P$=4000 nm along $x$ and $y$ axis. The width and length of the PDMS bars are $W$=600 nm and $L$= 4000 nm. The thicknesses of the PDMS layer and MgF$_2$ layer are $d_P$=750 nm and $d_M$=1500 nm. The electron doping of BP $n_s$=3$\times$10$^{13}$ cm$^{-2}$ is adapted. Under the normal incidence, the simulated absorption spectra for TM and TE polarizations are depicted in Fig. 2(a) and (b). It is observed that the absorption spectra show different properties dependent on polarization directions due to the in-plane anisotropy of BP. The spectrum for TM polarization shows a perfect absorption of 99.65$\%$ at the resonance wavelength of 4298.68 nm. Meanwhile, the spectrum for TE polarization exhibits absorption peak of 98.48$\%$ at 4308.98 nm and its absorption efficiency is only 2.61$\%$ at the resonance wavelength of TM polarization. 

In the proposed one-port system, the coupled mode theory (CMT) can be utilized to describe the resonance system, and the reflection coefficient is given by\cite{piper2014total}
\begin{equation}
\label{eq:1}
\Gamma=\frac{i(\omega-\omega_{0})+\delta-\gamma}{i(\omega-\omega_{0})+\delta+\gamma}.
\end{equation}
And the absorption is calculated from
\begin{equation}
\label{eq:2}
A=1-|\Gamma|^{2}=\frac{4\delta\gamma}{(\omega-\omega_{0})^{2}+(\delta+\gamma)^{2}},
\end{equation}
where $\omega$, $\delta$, $\gamma$ represents resonant frequency, intrinsic loss and external leakage rate, respectively. According to Eq. (2), the incident light will be completely absorption ($A$ =1) when the external leakage $\gamma$ equals the intrinsic loss $\delta$ at the resonance frequency ($\omega$=$\omega_0$). Under the critical coupling condition, the effective impedance of the whole structure $Z$ would be same with that of the free space $Z_0$, i.e. $Z$=$Z_0$=1, where the impedance $Z$ of the structure is derived from\cite{smith2005electromagnetic} 
\begin{equation}
\label{eq:3}
Z=\frac{(T_{22}-T_{11})\pm\sqrt{(T_{22}-T_{11})^{2}+4T_{12}T_{21}}}{2T_{21}}.
\end{equation}
Here $T_{11}$ $T_{12}$, $T_{21}$, $T_{22}$ are the elements of the transfer matrix calculated from the scattering matrix elements. 

The theoretical absorption spectra based on CMT are also depicted for TM and TE polarizations in Fig. 2(a) and (b), showing a good agreement with the simulated results. The absorption characteristic for TM polarization are analyzed because of the total absorption feature. The loss and external leakage of the structure from theoretical fitting are $\delta$=$\gamma$=1.03$\times$10$^{11}$ Hz. The quality factor Q is calculated as 1045.90 with the definition of $Q=\lambda_0/\Delta\lambda$, where the full width at half maximum (FWHM) is $\Delta\lambda=4.11$. Comparing with the theoretical $Q$ of 1064.31 by $Q_{CMT}=Q_{\delta}Q_{\gamma}/(Q_{\delta}+Q_{\gamma})$ from intrinsic loss $Q_{\delta}=\omega_0/(2\delta)$ and the external leakage $Q_{\gamma}=\omega_0/(2\gamma)$, the similar values between theoretical and simulated $Q$ illustrate the critical coupling mechanism for perfect absorption. In addition, the effective impedance of the structure is calculated as $Z = 1.03-i0.08$ at resonance wavelength, close to that of the free space. Hence the reflection of the structure is minimized through impedance matching with free space and the transmission is blocked by the metallic mirror, resulting in the perfect absorption of the entire structure. In comparison, for TE polarization, the intrinsic loss of the structure is calculated as 8.20$\times$10$^{10}$ Hz because of the smaller imaginary part of effective permittivity of BP along $y$ axis than that along $x$ axis, while the external leakage of the structure of the structure doesn’t change during the polarization shift. Meanwhile, the impedance is calculated as $Z=1.06+i0.21$ for the non-perfect absorption efficiency under the over coupling state. In addition, the electric filed distributions $|E|$ at the resonance wavelength of TM polarization are depicted in Fig. 2(c). Under the resonance state, the guided resonance confines the electric filed near the BP layer, leading to the significant enhancement of absorption of the structure. 

As mentioned above, the critically coupling mechanism of BP with guided resonance is employed to realize perfect absorption of the structure. The absorption characteristics of the structure can be actively controlled via manipulating the critical coupling conditions. Since BP is the main contributor to the loss of the structure in the wavelengths of interest, we take advantage of its tunable electron doping to tailor light absorption. Fig. 3 (a) and (b) illustrate the dependences of the absorption peaks and peak wavelengths on the electron doping of BP for TM and TE polarizations. When the electron doping of BP increases from 1$\times$10$^{13}$ cm$^{-2}$ to 3$\times$10$^{13}$ cm$^{-2}$, the absorption peak for TM polarization increases from 97.87$\%$ (4310.16 nm) to 99.65$\%$ (4298.86 nm) and then shows a decline tendency to 86.84$\%$ (4255.66 nm) with $n_s$=11$\times$10$^{13}$ cm$^{-2}$. During the variations, the imaginary part of effective permittivity of BP becomes larger, causing the increase of the intrinsic loss of the structure. The modulation process can also be theoretically explained based on CMT. The leakage rate $\gamma$ is considered as unchanged, $\gamma$= 1.03$\times$10$^{11}$. The loss of the structure is calculated as $\delta$=7.62$\times$10$^{10}$ Hz, 1.03$\times$10$^{11}$ Hz, and 2.19$\times$10$^{11}$ Hz for $n_s$= 1$\times$10$^{13}$ cm$^{-2}$, 3$\times$10$^{13}$ cm$^{-2}$ and 11$\times$10$^{13}$ cm$^{-2}$. In terms of the match degree of $\delta$ and $\gamma$, the system goes through the states from over coupling to critical coupling and under coupling with the increasing electron doping, which is reflected by the variations of absorption peaks in Fig. 3(a). For the same reason, the absorption peak for TE polarizations exhibits the similar tendency in Fig. 3(b). Meanwhile, because of the smaller real part of the effective permittivity of BP as $n_s$ increases, resonance wavelengths for both polarizations display a slight blue shift, indicating its feasibility within a wide range.

\begin{figure} [htbp]
	\label{figure3}
	\includegraphics
 	[scale=0.45]{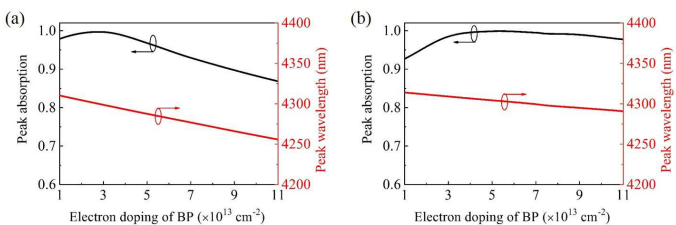} 
	\caption{\label{fig:FIG3}{Variations of absorption peaks and peak wavelengths with the increasing electron doping of BP for (a) TM and (b)TE polarizations.			
			}}	

\end{figure}

In addition to the modulation of intrinsic loss, the absorption characteristics of the structure can also be controlled by adjusting the leakage rate. Fig. 4 depicts the influence of the geometrical parameters on the absorption peaks and resonance wavelengths for TM polarization. When the width of the crossed PDMS bars increases with other parameters fixed as those in initial setting, the absorption peak gradually becomes larger from 64.48$\%$ (4253.65nm) for W=400 nm to 99.65$\%$ for W= 600nm and then decreases to 98.20$\%$ (4336.3nm) for W= 800 nm. This can also be explained by the match degree of the critical coupling conditions. The leakage rate $\gamma$ of the structure is theoretically calculated as 2.63$\times$10$^{10}$ Hz, 1.03$\times$10$^{11}$ Hz, and 1.33$\times$10$^{11}$ Hz for the three cases of W= 400 nm, 600 nm and 800 nm, respectively. Compared with the unchanged intrinsic loss of the whole structure, the system evolves from the states of under coupling, critical coupling to over coupling, resulting in the variations of the absorption peak. Also, due to the increasing effective refractive index, the resonance wavelength shows a red shift as the width of the PDMS bars increases. In Fig. 4 (b) and (c), the absorption characteristics show similar variation tendency with increasing thicknesses of PDMS layer and MgF$_2$ layer, but the absorption efficiency shows a more significant change, i.e. more sensitive than that to the width of PDMS bars. Hence, with a proper engineering of the geometrical parameters, the critical coupling conditions can be adjusted and the absorption characteristics of the structure will be flexibly tailored.
 
\begin{figure} [htbp]
	\label{figure4}
	\includegraphics
	[scale=0.45]{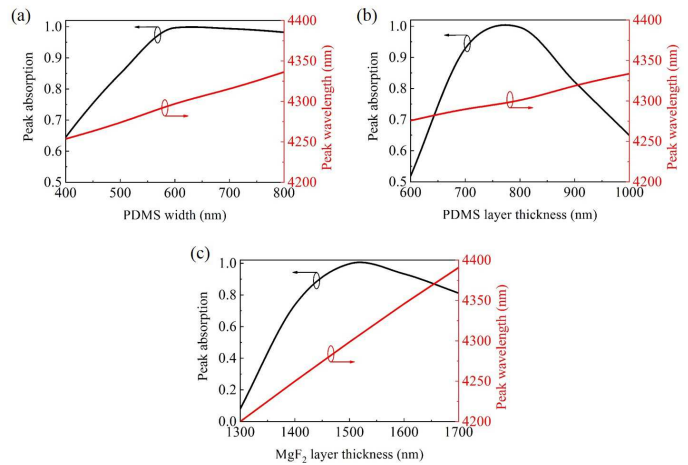} 
	\caption{\label{fig:FIG4}{Variations of absorption peaks and peak wavelengths for TM polarization with the geometrical parameters, including (a) width of the crossed PDMS bars, (b) thicknesses of PDMS layer and (c) thicknesses of MgF$_2$ layer. 
			}}	

\end{figure}

Then dependences of the absorption characteristics of the structure on different incident angles for TM and TE polarization are also investigated, as shown in Fig. 5. When the incident angles increase from 0$^{\circ}$ to 6$^{\circ}$, the absorption efficiency remains almost unchanged with the values above 94$\%$ due the insensitive properties of the guided resonance to the incident angles. It is also observed that the peak wavelength shows a slight blue shift from 4298.4 nm to 4281.23 nm for TM polarization. For TE polarization, two absorption peaks of the spectra are observed under different incident angles and the wavelength splitting in Fig.5 (b) arises from the excitation of the two guided resonances. The absorption characteristics for TE polarization under different incident angles will find its promising application in multispectral detection.

\begin{figure} [htbp]
	\label{figure5}
	\includegraphics
	[scale=0.45]{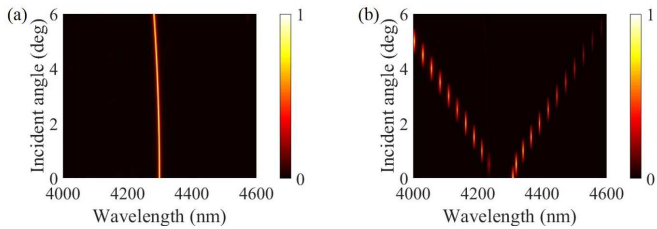} 
	\caption{\label{fig:FIG5}{Variations of absorption characteristics of the structure as the incident angles for (a) TM polarization and (b) TE polarization.		
			}}	

\end{figure}

In summary, a simple monolayer BP-based absorption metamaterial is proposed and the absorption characteristics of the structure are numerically simulated and theoretically analyzed. By critically coupling of the guided resonance, the absorption of the structure can be enhanced to 99.65$\%$ at resonance wavelength of 4298.68 nm for TM polarization while the absorption efficiency for TE polarization is only 2.61$\%$ at the same wavelength due to the anisotropic nature of BP. In addition to the high absorption efficiency, the structure also exhibits tunable absorption by changing the electron doping of the BP, geometrical parameters and incident angles. The proposed structure provides guidance for designing high performance BP-based devices with high absorption, remarkable anisotropy and flexible tunability.

\section*{acknowledgment}
This work is supported by the National Natural Science Foundation of China (Grant No. 11847132 and 61901164), Research Project of Hubei Provincial Department of Education (Q20193006) and Interdisciplinary Innovation Fund of Nanchang University (Grant No. 2019-9166-27060003).



%



\begin{thebibliography}{30}%
	\makeatletter
	\providecommand \@ifxundefined [1]{%
		\@ifx{#1\undefined}
	}%
	\providecommand \@ifnum [1]{%
		\ifnum #1\expandafter \@firstoftwo
		\else \expandafter \@secondoftwo
		\fi
	}%
	\providecommand \@ifx [1]{%
		\ifx #1\expandafter \@firstoftwo
		\else \expandafter \@secondoftwo
		\fi
	}%
	\providecommand \natexlab [1]{#1}%
	\providecommand \enquote  [1]{``#1''}%
	\providecommand \bibnamefont  [1]{#1}%
	\providecommand \bibfnamefont [1]{#1}%
	\providecommand \citenamefont [1]{#1}%
	\providecommand \href@noop [0]{\@secondoftwo}%
	\providecommand \href [0]{\begingroup \@sanitize@url \@href}%
	\providecommand \@href[1]{\@@startlink{#1}\@@href}%
	\providecommand \@@href[1]{\endgroup#1\@@endlink}%
	\providecommand \@sanitize@url [0]{\catcode `\\12\catcode `\$12\catcode
		`\&12\catcode `\#12\catcode `\^12\catcode `\_12\catcode `\%12\relax}%
	\providecommand \@@startlink[1]{}%
	\providecommand \@@endlink[0]{}%
	\providecommand \url  [0]{\begingroup\@sanitize@url \@url }%
	\providecommand \@url [1]{\endgroup\@href {#1}{\urlprefix }}%
	\providecommand \urlprefix  [0]{URL }%
	\providecommand \Eprint [0]{\href }%
	\providecommand \doibase [0]{http://dx.doi.org/}%
	\providecommand \selectlanguage [0]{\@gobble}%
	\providecommand \bibinfo  [0]{\@secondoftwo}%
	\providecommand \bibfield  [0]{\@secondoftwo}%
	\providecommand \translation [1]{[#1]}%
	\providecommand \BibitemOpen [0]{}%
	\providecommand \bibitemStop [0]{}%
	\providecommand \bibitemNoStop [0]{.\EOS\space}%
	\providecommand \EOS [0]{\spacefactor3000\relax}%
	\providecommand \BibitemShut  [1]{\csname bibitem#1\endcsname}%
	\let\auto@bib@innerbib\@empty
	\bibitem [{\citenamefont {Xia}\ \emph {et~al.}(2014)\citenamefont {Xia},
		\citenamefont {Wang}, \citenamefont {Xiao}, \citenamefont {Dubey},\ and\
		\citenamefont {Ramasubramaniam}}]{xia2014two}%
	\BibitemOpen
	\bibfield  {author} {\bibinfo {author} {\bibfnamefont {F.}~\bibnamefont
			{Xia}}, \bibinfo {author} {\bibfnamefont {H.}~\bibnamefont {Wang}}, \bibinfo
		{author} {\bibfnamefont {D.}~\bibnamefont {Xiao}}, \bibinfo {author}
		{\bibfnamefont {M.}~\bibnamefont {Dubey}}, \ and\ \bibinfo {author}
		{\bibfnamefont {A.}~\bibnamefont {Ramasubramaniam}},\ }\href@noop {}
	{\bibfield  {journal} {\bibinfo  {journal} {Nat. Photonics}\ }\textbf
		{\bibinfo {volume} {8}},\ \bibinfo {pages} {899} (\bibinfo {year}
		{2014})}\BibitemShut {NoStop}%
	\bibitem [{\citenamefont {Li}\ \emph {et~al.}(2014)\citenamefont {Li},
		\citenamefont {Yu}, \citenamefont {Ye}, \citenamefont {Ge}, \citenamefont
		{Ou}, \citenamefont {Wu}, \citenamefont {Feng}, \citenamefont {Chen},\ and\
		\citenamefont {Zhang}}]{li2014black}%
	\BibitemOpen
	\bibfield  {author} {\bibinfo {author} {\bibfnamefont {L.}~\bibnamefont
			{Li}}, \bibinfo {author} {\bibfnamefont {Y.}~\bibnamefont {Yu}}, \bibinfo
		{author} {\bibfnamefont {G.~J.}\ \bibnamefont {Ye}}, \bibinfo {author}
		{\bibfnamefont {Q.}~\bibnamefont {Ge}}, \bibinfo {author} {\bibfnamefont
			{X.}~\bibnamefont {Ou}}, \bibinfo {author} {\bibfnamefont {H.}~\bibnamefont
			{Wu}}, \bibinfo {author} {\bibfnamefont {D.}~\bibnamefont {Feng}}, \bibinfo
		{author} {\bibfnamefont {X.~H.}\ \bibnamefont {Chen}}, \ and\ \bibinfo
		{author} {\bibfnamefont {Y.}~\bibnamefont {Zhang}},\ }\href@noop {}
	{\bibfield  {journal} {\bibinfo  {journal} {Nat. Nanotechnol.}\ }\textbf
		{\bibinfo {volume} {9}},\ \bibinfo {pages} {372} (\bibinfo {year}
		{2014})}\BibitemShut {NoStop}%
	\bibitem [{\citenamefont {Xia}, \citenamefont {Wang},\ and\ \citenamefont
		{Jia}(2014)}]{xia2014rediscovering}%
	\BibitemOpen
	\bibfield  {author} {\bibinfo {author} {\bibfnamefont {F.}~\bibnamefont
			{Xia}}, \bibinfo {author} {\bibfnamefont {H.}~\bibnamefont {Wang}}, \ and\
		\bibinfo {author} {\bibfnamefont {Y.}~\bibnamefont {Jia}},\ }\href@noop {}
	{\bibfield  {journal} {\bibinfo  {journal} {Nat. Commun.}\ }\textbf {\bibinfo
			{volume} {5}},\ \bibinfo {pages} {4458} (\bibinfo {year} {2014})}\BibitemShut
	{NoStop}%
	\bibitem [{\citenamefont {Liu}\ \emph {et~al.}(2014)\citenamefont {Liu},
		\citenamefont {Neal}, \citenamefont {Zhu}, \citenamefont {Luo}, \citenamefont
		{Xu}, \citenamefont {Tom{\'a}nek},\ and\ \citenamefont
		{Ye}}]{liu2014phosphorene}%
	\BibitemOpen
	\bibfield  {author} {\bibinfo {author} {\bibfnamefont {H.}~\bibnamefont
			{Liu}}, \bibinfo {author} {\bibfnamefont {A.~T.}\ \bibnamefont {Neal}},
		\bibinfo {author} {\bibfnamefont {Z.}~\bibnamefont {Zhu}}, \bibinfo {author}
		{\bibfnamefont {Z.}~\bibnamefont {Luo}}, \bibinfo {author} {\bibfnamefont
			{X.}~\bibnamefont {Xu}}, \bibinfo {author} {\bibfnamefont {D.}~\bibnamefont
			{Tom{\'a}nek}}, \ and\ \bibinfo {author} {\bibfnamefont {P.~D.}\ \bibnamefont
			{Ye}},\ }\href@noop {} {\bibfield  {journal} {\bibinfo  {journal} {ACS Nano}\
		}\textbf {\bibinfo {volume} {8}},\ \bibinfo {pages} {4033} (\bibinfo {year}
		{2014})}\BibitemShut {NoStop}%
	\bibitem [{\citenamefont {Rodin}, \citenamefont {Carvalho},\ and\ \citenamefont
		{Neto}(2014)}]{rodin2014strain}%
	\BibitemOpen
	\bibfield  {author} {\bibinfo {author} {\bibfnamefont {A.}~\bibnamefont
			{Rodin}}, \bibinfo {author} {\bibfnamefont {A.}~\bibnamefont {Carvalho}}, \
		and\ \bibinfo {author} {\bibfnamefont {A.~C.}\ \bibnamefont {Neto}},\
	}\href@noop {} {\bibfield  {journal} {\bibinfo  {journal} {Phys. Rev. Lett.}\
		}\textbf {\bibinfo {volume} {112}},\ \bibinfo {pages} {176801} (\bibinfo
		{year} {2014})}\BibitemShut {NoStop}%
	\bibitem [{\citenamefont {Tran}\ \emph {et~al.}(2014)\citenamefont {Tran},
		\citenamefont {Soklaski}, \citenamefont {Liang},\ and\ \citenamefont
		{Yang}}]{tran2014layer}%
	\BibitemOpen
	\bibfield  {author} {\bibinfo {author} {\bibfnamefont {V.}~\bibnamefont
			{Tran}}, \bibinfo {author} {\bibfnamefont {R.}~\bibnamefont {Soklaski}},
		\bibinfo {author} {\bibfnamefont {Y.}~\bibnamefont {Liang}}, \ and\ \bibinfo
		{author} {\bibfnamefont {L.}~\bibnamefont {Yang}},\ }\href@noop {} {\bibfield
		{journal} {\bibinfo  {journal} {Phys. Rev. B}\ }\textbf {\bibinfo {volume}
			{89}},\ \bibinfo {pages} {235319} (\bibinfo {year} {2014})}\BibitemShut
	{NoStop}%
	\bibitem [{\citenamefont {Deng}\ \emph {et~al.}(2017)\citenamefont {Deng},
		\citenamefont {Tran}, \citenamefont {Xie}, \citenamefont {Jiang},
		\citenamefont {Li}, \citenamefont {Guo}, \citenamefont {Wang}, \citenamefont
		{Tian}, \citenamefont {Koester}, \citenamefont {Wang} \emph
		{et~al.}}]{deng2017efficient}%
	\BibitemOpen
	\bibfield  {author} {\bibinfo {author} {\bibfnamefont {B.}~\bibnamefont
			{Deng}}, \bibinfo {author} {\bibfnamefont {V.}~\bibnamefont {Tran}}, \bibinfo
		{author} {\bibfnamefont {Y.}~\bibnamefont {Xie}}, \bibinfo {author}
		{\bibfnamefont {H.}~\bibnamefont {Jiang}}, \bibinfo {author} {\bibfnamefont
			{C.}~\bibnamefont {Li}}, \bibinfo {author} {\bibfnamefont {Q.}~\bibnamefont
			{Guo}}, \bibinfo {author} {\bibfnamefont {X.}~\bibnamefont {Wang}}, \bibinfo
		{author} {\bibfnamefont {H.}~\bibnamefont {Tian}}, \bibinfo {author}
		{\bibfnamefont {S.~J.}\ \bibnamefont {Koester}}, \bibinfo {author}
		{\bibfnamefont {H.}~\bibnamefont {Wang}},  \emph {et~al.},\ }\href@noop {}
	{\bibfield  {journal} {\bibinfo  {journal} {Nat. Commun.}\ }\textbf {\bibinfo
			{volume} {8}},\ \bibinfo {pages} {14474} (\bibinfo {year}
		{2017})}\BibitemShut {NoStop}%
	\bibitem [{\citenamefont {Fei}\ and\ \citenamefont
		{Yang}(2014)}]{fei2014strain}%
	\BibitemOpen
	\bibfield  {author} {\bibinfo {author} {\bibfnamefont {R.}~\bibnamefont
			{Fei}}\ and\ \bibinfo {author} {\bibfnamefont {L.}~\bibnamefont {Yang}},\
	}\href@noop {} {\bibfield  {journal} {\bibinfo  {journal} {Nano Lett.}\
		}\textbf {\bibinfo {volume} {14}},\ \bibinfo {pages} {2884} (\bibinfo {year}
		{2014})}\BibitemShut {NoStop}%
	\bibitem [{\citenamefont {Qiao}\ \emph {et~al.}(2014)\citenamefont {Qiao},
		\citenamefont {Kong}, \citenamefont {Hu}, \citenamefont {Yang},\ and\
		\citenamefont {Ji}}]{qiao2014high}%
	\BibitemOpen
	\bibfield  {author} {\bibinfo {author} {\bibfnamefont {J.}~\bibnamefont
			{Qiao}}, \bibinfo {author} {\bibfnamefont {X.}~\bibnamefont {Kong}}, \bibinfo
		{author} {\bibfnamefont {Z.-X.}\ \bibnamefont {Hu}}, \bibinfo {author}
		{\bibfnamefont {F.}~\bibnamefont {Yang}}, \ and\ \bibinfo {author}
		{\bibfnamefont {W.}~\bibnamefont {Ji}},\ }\href@noop {} {\bibfield  {journal}
		{\bibinfo  {journal} {Nat. Commun.}\ }\textbf {\bibinfo {volume} {5}},\
		\bibinfo {pages} {4475} (\bibinfo {year} {2014})}\BibitemShut {NoStop}%
	\bibitem [{\citenamefont {Low}\ \emph {et~al.}(2014)\citenamefont {Low},
		\citenamefont {Rold{\'a}n}, \citenamefont {Wang}, \citenamefont {Xia},
		\citenamefont {Avouris}, \citenamefont {Moreno},\ and\ \citenamefont
		{Guinea}}]{low2014plasmons}%
	\BibitemOpen
	\bibfield  {author} {\bibinfo {author} {\bibfnamefont {T.}~\bibnamefont
			{Low}}, \bibinfo {author} {\bibfnamefont {R.}~\bibnamefont {Rold{\'a}n}},
		\bibinfo {author} {\bibfnamefont {H.}~\bibnamefont {Wang}}, \bibinfo {author}
		{\bibfnamefont {F.}~\bibnamefont {Xia}}, \bibinfo {author} {\bibfnamefont
			{P.}~\bibnamefont {Avouris}}, \bibinfo {author} {\bibfnamefont {L.~M.}\
			\bibnamefont {Moreno}}, \ and\ \bibinfo {author} {\bibfnamefont
			{F.}~\bibnamefont {Guinea}},\ }\href@noop {} {\bibfield  {journal} {\bibinfo
			{journal} {Phys. Rev. Lett.}\ }\textbf {\bibinfo {volume} {113}},\ \bibinfo
		{pages} {106802} (\bibinfo {year} {2014})}\BibitemShut {NoStop}%
	\bibitem [{\citenamefont {Liu}\ and\ \citenamefont
		{Aydin}(2016)}]{liu2016localized}%
	\BibitemOpen
	\bibfield  {author} {\bibinfo {author} {\bibfnamefont {Z.}~\bibnamefont
			{Liu}}\ and\ \bibinfo {author} {\bibfnamefont {K.}~\bibnamefont {Aydin}},\
	}\href@noop {} {\bibfield  {journal} {\bibinfo  {journal} {Nano Lett.}\
		}\textbf {\bibinfo {volume} {16}},\ \bibinfo {pages} {3457} (\bibinfo {year}
		{2016})}\BibitemShut {NoStop}%
	\bibitem [{\citenamefont {Lu}\ \emph {et~al.}(2017)\citenamefont {Lu},
		\citenamefont {Gong}, \citenamefont {Mao}, \citenamefont {Gan},\ and\
		\citenamefont {Zhao}}]{lu2017strong}%
	\BibitemOpen
	\bibfield  {author} {\bibinfo {author} {\bibfnamefont {H.}~\bibnamefont
			{Lu}}, \bibinfo {author} {\bibfnamefont {Y.}~\bibnamefont {Gong}}, \bibinfo
		{author} {\bibfnamefont {D.}~\bibnamefont {Mao}}, \bibinfo {author}
		{\bibfnamefont {X.}~\bibnamefont {Gan}}, \ and\ \bibinfo {author}
		{\bibfnamefont {J.}~\bibnamefont {Zhao}},\ }\href@noop {} {\bibfield
		{journal} {\bibinfo  {journal} {Opt. Express}\ }\textbf {\bibinfo {volume}
			{25}},\ \bibinfo {pages} {5255} (\bibinfo {year} {2017})}\BibitemShut
	{NoStop}%
	\bibitem [{\citenamefont {Wang}, \citenamefont {Jiang},\ and\ \citenamefont
		{Hu}(2017)}]{wang2017dual}%
	\BibitemOpen
	\bibfield  {author} {\bibinfo {author} {\bibfnamefont {J.}~\bibnamefont
			{Wang}}, \bibinfo {author} {\bibfnamefont {Y.}~\bibnamefont {Jiang}}, \ and\
		\bibinfo {author} {\bibfnamefont {Z.}~\bibnamefont {Hu}},\ }\href@noop {}
	{\bibfield  {journal} {\bibinfo  {journal} {Opt. Express}\ }\textbf {\bibinfo
			{volume} {25}},\ \bibinfo {pages} {22149} (\bibinfo {year}
		{2017})}\BibitemShut {NoStop}%
	\bibitem [{\citenamefont {Song}\ \emph {et~al.}(2018)\citenamefont {Song},
		\citenamefont {Liu}, \citenamefont {Xiang},\ and\ \citenamefont
		{Aydin}}]{song2018biaxial}%
	\BibitemOpen
	\bibfield  {author} {\bibinfo {author} {\bibfnamefont {X.}~\bibnamefont
			{Song}}, \bibinfo {author} {\bibfnamefont {Z.}~\bibnamefont {Liu}}, \bibinfo
		{author} {\bibfnamefont {Y.}~\bibnamefont {Xiang}}, \ and\ \bibinfo {author}
		{\bibfnamefont {K.}~\bibnamefont {Aydin}},\ }\href@noop {} {\bibfield
		{journal} {\bibinfo  {journal} {Opt. Express}\ }\textbf {\bibinfo {volume}
			{26}},\ \bibinfo {pages} {5469} (\bibinfo {year} {2018})}\BibitemShut
	{NoStop}%
	\bibitem [{\citenamefont {Fo}\ \emph {et~al.}(2018)\citenamefont {Fo},
		\citenamefont {Pan}, \citenamefont {Chen}, \citenamefont {Xu}, \citenamefont
		{Ouyang}, \citenamefont {Zhang}, \citenamefont {Tian}, \citenamefont {Gu},
		\citenamefont {Liu}, \citenamefont {Han} \emph {et~al.}}]{fo2018anisotropic}%
	\BibitemOpen
	\bibfield  {author} {\bibinfo {author} {\bibfnamefont {Q.}~\bibnamefont
			{Fo}}, \bibinfo {author} {\bibfnamefont {L.}~\bibnamefont {Pan}}, \bibinfo
		{author} {\bibfnamefont {X.}~\bibnamefont {Chen}}, \bibinfo {author}
		{\bibfnamefont {Q.}~\bibnamefont {Xu}}, \bibinfo {author} {\bibfnamefont
			{C.}~\bibnamefont {Ouyang}}, \bibinfo {author} {\bibfnamefont
			{X.}~\bibnamefont {Zhang}}, \bibinfo {author} {\bibfnamefont
			{Z.}~\bibnamefont {Tian}}, \bibinfo {author} {\bibfnamefont {J.}~\bibnamefont
			{Gu}}, \bibinfo {author} {\bibfnamefont {L.}~\bibnamefont {Liu}}, \bibinfo
		{author} {\bibfnamefont {J.}~\bibnamefont {Han}},  \emph {et~al.},\
	}\href@noop {} {\bibfield  {journal} {\bibinfo  {journal} {IEEE Photon. J.}\
		}\textbf {\bibinfo {volume} {10}},\ \bibinfo {pages} {1} (\bibinfo {year}
		{2018})}\BibitemShut {NoStop}%
	\bibitem [{\citenamefont {Guo}\ and\ \citenamefont
		{Argyropoulos}(2019)}]{guo2019tunable}%
	\BibitemOpen
	\bibfield  {author} {\bibinfo {author} {\bibfnamefont {T.}~\bibnamefont
			{Guo}}\ and\ \bibinfo {author} {\bibfnamefont {C.}~\bibnamefont
			{Argyropoulos}},\ }\href@noop {} {\bibfield  {journal} {\bibinfo  {journal}
			{JOSA B}\ }\textbf {\bibinfo {volume} {36}},\ \bibinfo {pages} {2962}
		(\bibinfo {year} {2019})}\BibitemShut {NoStop}%
	\bibitem [{\citenamefont {Xiao}\ \emph {et~al.}(2019)\citenamefont {Xiao},
		\citenamefont {Liu}, \citenamefont {Cheng}, \citenamefont {Zhou},
		\citenamefont {Jiang}, \citenamefont {Li},\ and\ \citenamefont
		{Xu}}]{xiao2019tunable}%
	\BibitemOpen
	\bibfield  {author} {\bibinfo {author} {\bibfnamefont {S.}~\bibnamefont
			{Xiao}}, \bibinfo {author} {\bibfnamefont {T.}~\bibnamefont {Liu}}, \bibinfo
		{author} {\bibfnamefont {L.}~\bibnamefont {Cheng}}, \bibinfo {author}
		{\bibfnamefont {C.}~\bibnamefont {Zhou}}, \bibinfo {author} {\bibfnamefont
			{X.}~\bibnamefont {Jiang}}, \bibinfo {author} {\bibfnamefont
			{Z.}~\bibnamefont {Li}}, \ and\ \bibinfo {author} {\bibfnamefont
			{C.}~\bibnamefont {Xu}},\ }\href@noop {} {\bibfield  {journal} {\bibinfo
			{journal} {J. Lightw. Technol.}\ } (\bibinfo {year} {2019})}\BibitemShut
	{NoStop}%
	\bibitem [{\citenamefont {Zhu}, \citenamefont {Tang},\ and\ \citenamefont
		{Jiang}(2019)}]{zhu2019tunable}%
	\BibitemOpen
	\bibfield  {author} {\bibinfo {author} {\bibfnamefont {Y.}~\bibnamefont
			{Zhu}}, \bibinfo {author} {\bibfnamefont {B.}~\bibnamefont {Tang}}, \ and\
		\bibinfo {author} {\bibfnamefont {C.}~\bibnamefont {Jiang}},\ }\href@noop {}
	{\bibfield  {journal} {\bibinfo  {journal} {Appl. Phys Express}\ }\textbf
		{\bibinfo {volume} {12}},\ \bibinfo {pages} {032009} (\bibinfo {year}
		{2019})}\BibitemShut {NoStop}%
	\bibitem [{\citenamefont {Cai}\ \emph {et~al.}(2019)\citenamefont {Cai},
		\citenamefont {Xu}, \citenamefont {Feng}, \citenamefont {Guo}, \citenamefont
		{Lin},\ and\ \citenamefont {Zhu}}]{cai2019anisotropic}%
	\BibitemOpen
	\bibfield  {author} {\bibinfo {author} {\bibfnamefont {Y.}~\bibnamefont
			{Cai}}, \bibinfo {author} {\bibfnamefont {K.-D.}\ \bibnamefont {Xu}},
		\bibinfo {author} {\bibfnamefont {N.}~\bibnamefont {Feng}}, \bibinfo {author}
		{\bibfnamefont {R.}~\bibnamefont {Guo}}, \bibinfo {author} {\bibfnamefont
			{H.}~\bibnamefont {Lin}}, \ and\ \bibinfo {author} {\bibfnamefont
			{J.}~\bibnamefont {Zhu}},\ }\href@noop {} {\bibfield  {journal} {\bibinfo
			{journal} {Opt. Express}\ }\textbf {\bibinfo {volume} {27}},\ \bibinfo
		{pages} {3101} (\bibinfo {year} {2019})}\BibitemShut {NoStop}%
	\bibitem [{\citenamefont {Piper}\ and\ \citenamefont
		{Fan}(2014)}]{piper2014total}%
	\BibitemOpen
	\bibfield  {author} {\bibinfo {author} {\bibfnamefont {J.~R.}\ \bibnamefont
			{Piper}}\ and\ \bibinfo {author} {\bibfnamefont {S.}~\bibnamefont {Fan}},\
	}\href@noop {} {\bibfield  {journal} {\bibinfo  {journal} {ACS Photonics}\
		}\textbf {\bibinfo {volume} {1}},\ \bibinfo {pages} {347} (\bibinfo {year}
		{2014})}\BibitemShut {NoStop}%
	\bibitem [{\citenamefont {Zand}\ \emph {et~al.}(2018)\citenamefont {Zand},
		\citenamefont {Dalir}, \citenamefont {Chen},\ and\ \citenamefont
		{Dowling}}]{zand2018multispectral}%
	\BibitemOpen
	\bibfield  {author} {\bibinfo {author} {\bibfnamefont {I.}~\bibnamefont
			{Zand}}, \bibinfo {author} {\bibfnamefont {H.}~\bibnamefont {Dalir}},
		\bibinfo {author} {\bibfnamefont {R.~T.}\ \bibnamefont {Chen}}, \ and\
		\bibinfo {author} {\bibfnamefont {J.~P.}\ \bibnamefont {Dowling}},\
	}\href@noop {} {\bibfield  {journal} {\bibinfo  {journal} {Appl. Phys
				Express}\ }\textbf {\bibinfo {volume} {11}},\ \bibinfo {pages} {035101}
		(\bibinfo {year} {2018})}\BibitemShut {NoStop}%
	\bibitem [{\citenamefont {Luo}\ \emph {et~al.}(2018)\citenamefont {Luo},
		\citenamefont {Liu}, \citenamefont {Wang}, \citenamefont {Liu},\ and\
		\citenamefont {Lin}}]{luo2018tunable}%
	\BibitemOpen
	\bibfield  {author} {\bibinfo {author} {\bibfnamefont {X.}~\bibnamefont
			{Luo}}, \bibinfo {author} {\bibfnamefont {Z.}~\bibnamefont {Liu}}, \bibinfo
		{author} {\bibfnamefont {L.}~\bibnamefont {Wang}}, \bibinfo {author}
		{\bibfnamefont {J.}~\bibnamefont {Liu}}, \ and\ \bibinfo {author}
		{\bibfnamefont {Q.}~\bibnamefont {Lin}},\ }\href@noop {} {\bibfield
		{journal} {\bibinfo  {journal} {Appl. Phys. Express}\ }\textbf {\bibinfo
			{volume} {11}},\ \bibinfo {pages} {105102} (\bibinfo {year}
		{2018})}\BibitemShut {NoStop}%
	\bibitem [{\citenamefont {Jiang}\ \emph {et~al.}(2018)\citenamefont {Jiang},
		\citenamefont {Wang}, \citenamefont {Xiao}, \citenamefont {Yan},
		\citenamefont {Cheng},\ and\ \citenamefont {Zhong}}]{jiang2018approaching}%
	\BibitemOpen
	\bibfield  {author} {\bibinfo {author} {\bibfnamefont {X.}~\bibnamefont
			{Jiang}}, \bibinfo {author} {\bibfnamefont {T.}~\bibnamefont {Wang}},
		\bibinfo {author} {\bibfnamefont {S.}~\bibnamefont {Xiao}}, \bibinfo {author}
		{\bibfnamefont {X.}~\bibnamefont {Yan}}, \bibinfo {author} {\bibfnamefont
			{L.}~\bibnamefont {Cheng}}, \ and\ \bibinfo {author} {\bibfnamefont
			{Q.}~\bibnamefont {Zhong}},\ }\href@noop {} {\bibfield  {journal} {\bibinfo
			{journal} {Nanotechnology}\ }\textbf {\bibinfo {volume} {29}},\ \bibinfo
		{pages} {335205} (\bibinfo {year} {2018})}\BibitemShut {NoStop}%
	\bibitem [{\citenamefont {Li}\ \emph {et~al.}(2018)\citenamefont {Li},
		\citenamefont {Ren}, \citenamefont {Hu}, \citenamefont {Qin},\ and\
		\citenamefont {Wang}}]{li2018wavelength}%
	\BibitemOpen
	\bibfield  {author} {\bibinfo {author} {\bibfnamefont {H.-J.}\ \bibnamefont
			{Li}}, \bibinfo {author} {\bibfnamefont {Y.-Z.}\ \bibnamefont {Ren}},
		\bibinfo {author} {\bibfnamefont {J.-G.}\ \bibnamefont {Hu}}, \bibinfo
		{author} {\bibfnamefont {M.}~\bibnamefont {Qin}}, \ and\ \bibinfo {author}
		{\bibfnamefont {L.-L.}\ \bibnamefont {Wang}},\ }\href@noop {} {\bibfield
		{journal} {\bibinfo  {journal} {J. Lightw. Technol.}\ }\textbf {\bibinfo
			{volume} {36}},\ \bibinfo {pages} {3236} (\bibinfo {year}
		{2018})}\BibitemShut {NoStop}%
	\bibitem [{\citenamefont {Qing}, \citenamefont {Ma},\ and\ \citenamefont
		{Cui}(2018)}]{qing2018tailoring}%
	\BibitemOpen
	\bibfield  {author} {\bibinfo {author} {\bibfnamefont {Y.~M.}\ \bibnamefont
			{Qing}}, \bibinfo {author} {\bibfnamefont {H.~F.}\ \bibnamefont {Ma}}, \ and\
		\bibinfo {author} {\bibfnamefont {T.~J.}\ \bibnamefont {Cui}},\ }\href@noop
	{} {\bibfield  {journal} {\bibinfo  {journal} {Opt. Express}\ }\textbf
		{\bibinfo {volume} {26}},\ \bibinfo {pages} {32442} (\bibinfo {year}
		{2018})}\BibitemShut {NoStop}%
	\bibitem [{\citenamefont {Dodge}(1984)}]{dodge1984refractive}%
	\BibitemOpen
	\bibfield  {author} {\bibinfo {author} {\bibfnamefont {M.~J.}\ \bibnamefont
			{Dodge}},\ }\href@noop {} {\bibfield  {journal} {\bibinfo  {journal} {Appl.
				Opt.}\ }\textbf {\bibinfo {volume} {23}},\ \bibinfo {pages} {1980} (\bibinfo
		{year} {1984})}\BibitemShut {NoStop}%
	\bibitem [{\citenamefont {Querry}(1987)}]{querry1987optical}%
	\BibitemOpen
	\bibfield  {author} {\bibinfo {author} {\bibfnamefont {M.}~\bibnamefont
			{Querry}},\ }\href@noop {} {\enquote {\bibinfo {title} {Optical constants of
				minerals and other materials from the millimeter to the ultraviolet},}\
	}\bibinfo {type} {Tech. Rep.}\ (\bibinfo  {institution} {Chemical Research
		Development And Engineering Center Aberdeen Proving Groundmd},\ \bibinfo
	{year} {1987})\BibitemShut {NoStop}%
	\bibitem [{\citenamefont {Xiong}\ \emph {et~al.}(2017)\citenamefont {Xiong},
		\citenamefont {Zhang}, \citenamefont {Zhu}, \citenamefont {Yuan},\ and\
		\citenamefont {Qin}}]{xiong2017strong}%
	\BibitemOpen
	\bibfield  {author} {\bibinfo {author} {\bibfnamefont {F.}~\bibnamefont
			{Xiong}}, \bibinfo {author} {\bibfnamefont {J.}~\bibnamefont {Zhang}},
		\bibinfo {author} {\bibfnamefont {Z.}~\bibnamefont {Zhu}}, \bibinfo {author}
		{\bibfnamefont {X.}~\bibnamefont {Yuan}}, \ and\ \bibinfo {author}
		{\bibfnamefont {S.}~\bibnamefont {Qin}},\ }\href@noop {} {\bibfield
		{journal} {\bibinfo  {journal} {J. Opt.}\ }\textbf {\bibinfo {volume} {19}},\
		\bibinfo {pages} {075002} (\bibinfo {year} {2017})}\BibitemShut {NoStop}%
	\bibitem [{\citenamefont {Ordal}\ \emph {et~al.}(1985)\citenamefont {Ordal},
		\citenamefont {Bell}, \citenamefont {Alexander}, \citenamefont {Long},\ and\
		\citenamefont {Querry}}]{ordal1985optical}%
	\BibitemOpen
	\bibfield  {author} {\bibinfo {author} {\bibfnamefont {M.~A.}\ \bibnamefont
			{Ordal}}, \bibinfo {author} {\bibfnamefont {R.~J.}\ \bibnamefont {Bell}},
		\bibinfo {author} {\bibfnamefont {R.~W.}\ \bibnamefont {Alexander}}, \bibinfo
		{author} {\bibfnamefont {L.~L.}\ \bibnamefont {Long}}, \ and\ \bibinfo
		{author} {\bibfnamefont {M.~R.}\ \bibnamefont {Querry}},\ }\href@noop {}
	{\bibfield  {journal} {\bibinfo  {journal} {Appl. Opt.}\ }\textbf {\bibinfo
			{volume} {24}},\ \bibinfo {pages} {4493} (\bibinfo {year}
		{1985})}\BibitemShut {NoStop}%
	\bibitem [{\citenamefont {Smith}\ \emph {et~al.}(2005)\citenamefont {Smith},
		\citenamefont {Vier}, \citenamefont {Koschny},\ and\ \citenamefont
		{Soukoulis}}]{smith2005electromagnetic}%
	\BibitemOpen
	\bibfield  {author} {\bibinfo {author} {\bibfnamefont {D.}~\bibnamefont
			{Smith}}, \bibinfo {author} {\bibfnamefont {D.}~\bibnamefont {Vier}},
		\bibinfo {author} {\bibfnamefont {T.}~\bibnamefont {Koschny}}, \ and\
		\bibinfo {author} {\bibfnamefont {C.}~\bibnamefont {Soukoulis}},\ }\href@noop
	{} {\bibfield  {journal} {\bibinfo  {journal} {Phys. Rev. E}\ }\textbf
		{\bibinfo {volume} {71}},\ \bibinfo {pages} {036617} (\bibinfo {year}
		{2005})}\BibitemShut {NoStop}%
\end{thebibliography}

\end{document}